\documentstyle[prd,aps,preprint,epsf,epsfig]{revtex}
\firstfigfalse

\begin{document}

\draft

\preprint{\rightline{ANL-HEP-PR-01-027}}

\newcommand{\Dirac}{\rlap{\hspace{-.5mm} \slash} D}

\title{ 
 Diquark Condensation at Nonzero Chemical Potential and Temperature}
\vskip -1 truecm

\author{John~B.~Kogut and Dominique Toublan }
\address {Physics Department, University of Illinois at Urbana-Champaign,
Urbana, IL 61801-30}

\author{D.~K.~Sinclair.}
\address{HEP Division, Argonne National Laboratory, 9700
        South Cass Avenue, Argonne, IL 60439, USA}

\date{\today}
\maketitle
\vskip -1 truecm

\begin{abstract}

$SU(2)$ lattice gauge theory with four flavors of quarks is studied at nonzero chemical potential
$\mu$ and temperature $T$ by computer simulation and Effective Lagrangian techniques.
Simulations are done on $8^4$, $8^3 \times 4$
and $12^3 \times 6$ lattices and the diquark condensate,
chiral order parameter, Wilson line, fermion energy and number densities are measured.
Simulations at a fixed, nonzero quark mass provide evidence for a tricritical point in
the $\mu$-$T$ plane associated with diquark condensation. For low $T$, increasing $\mu$ takes
the system through a line of second order phase transitions to a diquark condensed phase.
Increasing $T$ at high $\mu$, the system passes through a line of first order transitions
from the diquark phase to the quark-gluon plasma phase. Using Effective Lagrangians we estimate
the position of the tricritical point and ascribe its existence to
trilinear couplings that increase with $\mu$ and $T$.

\end{abstract}

\newpage

Recently there has been a resurgence of interest in QCD at nonzero chemical potential
for quark number. Arguments
based on instantons \cite{Shuryak} and phenomenological gap equations
\cite{Wilczek} support the old idea that diquark condensation and a color
superconductivity phase transition \cite{Love} will occur at a critical chemical potential of
slightly less than one-third the proton's mass. Unfortunately, some of these
arguments and calculations could prove to be
misleading because a proper understanding of the diquark state alludes the field :
since diquarks carry color in real QCD, a quantitative understanding of confinement and
screening is needed to estimate the energies of the states which control the phases of the system. In
addition, the brute force lattice simulation method does not yet have a reliable simulation
algorithm for these environments in which the fermion determinant becomes complex.

Since true QCD at finite quark-number chemical potential cannot be simulated
with current methods, theorists have turned to simpler models which can be
simulated. One of the more interesting is the color $SU(2)$ version of QCD which addresses some of
the issues of interest \cite{Shuryak}, \cite{Wilczek}, \cite{Hands}. In this model
diquarks do not carry color, so their condensation does
not break color symmetry dynamically. The critical chemical potential is also expected to be
one-half the mass of the lightest meson, the pion, because quarks and
anti-quarks reside in equivalent representations of the $SU(2)$ color group. Chiral
Lagrangians can be used to study the diquark condensation transition in this model because the
critical chemical potential vanishes in the chiral limit, and the model has a Goldstone
realization of the spontaneously broken quark-number symmetry. Lattice simulations of the model are
also possible because the fermion determinant is non-negative for all chemical potentials. One
hopes that these developments will uncover generic phenomena that will also apply to
QCD at nonzero chemical potential.

Preliminary lattice simulations of the $SU(2)$ model with four species of quarks have been
presented at conferences recently \cite{DK2000}, and it is the purpose of this article to present
the continuation of that work with a focus on the system's phase diagram at nonzero $\mu$ and $T$.
Earlier work on
this model at finite $T$ and $\mu$ was performed by \cite{t+mu}. More detailed
discussion of our zero temperature, finite $\mu$ simulations including the
spectroscopy of the light bosonic modes will be presented elsewhere
\cite{inprep}.
We choose to consider the
system's phase diagram here because of its elementary, fundamental character and because these
simulations produced an unanticipated result compared to what has been found
previously in various models \cite{Dagotto,Benoit}. For fixed quark mass, which insures us that
the pion has a nonzero mass and chiral symmetry is explicitly broken, we find a line of
transitions separating a phase with no diquark condensation from one with a diquark condensate.
Along this line there is a tricritical point where the transition switches
from being second order (and well described by mean field theory) at relatively
low $\mu$ to a first order
transition at an intermediate $\mu$ value.
At large $\mu$ the line of first order transitions separates the diquark condensate phase
at low $T$ from a quark-gluon phase at high $T$. The transition is clear in diquark observables
as well as the Wilson Line. A schematic phase diagram is shown in Fig.1. (Note
that we have not included the line of first order transitions starting
from the $\mu=0$, finite temperature transition, which exists for $m$ small
enough).
We shall see that this simulation result, the existence of a tricritical point, has
a natural explanation in the context of chiral Lagrangians. Following the formalism of
\cite{Toublan} we shall argue that trilinear couplings among the low lying boson fields of
the Lagrangian become more significant as $\mu$ and $T$ increase and they can cause the
transition to become first order at a $\mu$ value in the vicinity of the results found in
the simulation.  By hindsight, this behaviour should not
have come as a surprise. $\mu$ plays the r\^{o}le of a second `temperature' in
this theory in that it is a parameter that controls diquark condensation, but
does not explicitly break the (quark-number) symmetry. The existence of two
competing `temperatures' is a characteristic of systems which exhibit
tricritical behaviour.

\begin{figure}

\centerline{
\epsfxsize 5 in
\epsfysize 3 in
\epsfbox{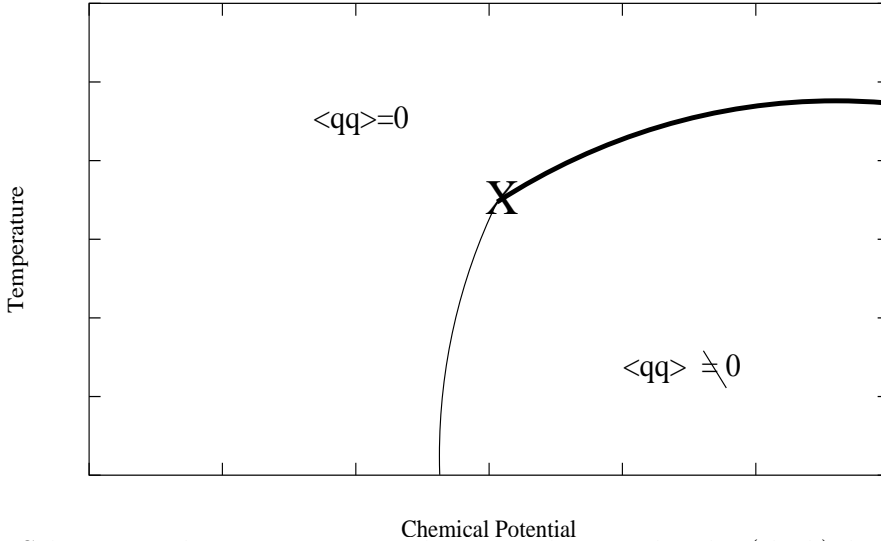}
}
\caption{Schematic Phase Diagram in the $T$-$\mu$ Plane. The thin(thick) line
consists of second(first) order transitions. X labels the tricritical point.}
\end{figure}

We begin with a discussion of the simulation method and the numerical results.

The lattice action of the staggered fermion version of this theory is:

\begin{equation}
S_f = \sum_{sites}\left\{\bar{\chi}[D\!\!\!\!/\,(\mu) + m]\chi
+ \frac{1}{2}\lambda[\chi^T\tau_2\chi + \bar{\chi}\tau_2\bar{\chi}^T]\right\}
\end{equation}

\noindent where the chemical potential $\mu$ is introduced by multiplying links in the
$+t$ direction by $e^\mu$ and those in the $-t$ direction by $e^{-\mu}$ \cite{First}. The
diquark source term (Majorana mass term) is added to allow us to observe
spontaneous breakdown of quark-number on a finite lattice. The parameter $\lambda$
and the usual mass term $m$ control the amount of explicit symmetry breaking in the
lattice action. We will be particularly interested in extrapolations to $\lambda \rightarrow 0$
for a given $m$ to produce an interesting, realistic
physical situation. This is the case that has been studied analytically using
effective Lagrangians at vanishing temperature $T$ but nonvanishing chemical potential $\mu$
\cite{Toublan}, \cite{SUNY}.

Integrating out the fermion fields in Eq.1 gives:

\begin{equation}
pfaffian\left[\begin{array}{cc} \lambda\tau_2     &    {\cal A}       \\
                                     -{\cal A}^T        &    \lambda\tau_2
\end{array}\right] = \sqrt{{\rm det}({\cal A}^\dagger {\cal A} + \lambda^2)}
\end{equation}
where
\begin{equation}
           {\cal A} \equiv  D\!\!\!\!/\,(\mu)+m
\end{equation}

Note that the pfaffian is strictly positive, so that we can use the hybrid
molecular dynamics \cite{HMD} method to simulate this theory using ``noisy'' fermions to
take the square root, giving $N_f=4$.

For $\lambda = 0$, $m \ne 0$, $\mu \ne 0$ we expect no spontaneous
symmetry breaking for small $\mu$. For $\mu$ large enough ($\mu > m_\pi/2$ according
to most approaches including Chiral Perturbation Theory \cite{Toublan}, \cite{SUNY})
we expect spontaneous breakdown of quark number and one Goldstone boson -- a
scalar diquark. (The reader should consult \cite{Hands}  for a full discussion of the symmetries
of the lattice action, remarks about spectroscopy, Goldstone as well as pseudo-Goldstone
bosons, and
for early simulations of the 8 flavour theory at $\lambda=0$.)

Now consider the simulation results for the $N_f=4$ theory
on $8^4$, $8^3 \times 4$ and $12^3 \times 6$ lattices, measuring the chiral
and diquark condensates ($\langle\chi^T\tau_2\chi\rangle$), the fermion number
density, the Wilson/Polyakov line,
etc. We are also engaged in larger scale simulations on $12^3 \times 24$ lattices, where,
in addition, we are measuring all local scalar and pseudoscalar meson and
diquark propagators (connected and disconnected). These simulations will
be presented in our paper on zero temperature results \cite{inprep}. Some of these results, especially for relatively large
quark mass have been previewed elsewhere \cite{DK2000}.

First consider measurements on a $8^4$, 'zero temperature' lattice. We simulated the $SU(2)$
model at a relatively strong coupling $\beta=1.0$ to avoid finite size and temperature effects.
(More extensive simulations at $\beta=1.5$ and $m=0.1$ also on an $8^4$ lattice
will be presented in our zero temperature paper.)
The quark mass was $m= 0.05$ and a series of simulations were done at
$\lambda = 0.0025$, $0.005$, and $0.01$ so that our results could be extrapolated to
vanishing diquark source, $\lambda=0$. Simple linear and spline extrapolations were done
and these procedures appeared to be sensible for these exploratory calculations. In the future
more theoretical control over the $\lambda \rightarrow 0$ limit will have to be developed.
None of the conclusions to be drawn here will depend strongly on the limit. In fact, everything we
say could be gathered from our data at a fixed (small) $\lambda$ value, $0.005$ say. However, only in the
limit of vanishing $\lambda$ do we expect real diquark phase transitions
(at least when the transitions are second order),
so it is interesting and relevant to begin investigating the  $\lambda \rightarrow 0$ limit.

Two further difficulties occur for small $\lambda$. First, the Hybrid algorithm slows down
as $\lambda$ is taken small and diquark observables are calculated. Analogous problems are
familiar in QCD simulations at vanishing $\mu$ but small $m$ because the Dirac
operator of the standard lattice action becomes singular in the chiral limit \cite{HMD}.
Even these exploratory
simulations are quite CPU intensive for this reason, although they are well within
the capabilities of standard PC's, and workstations. (Most of the results reported here
were done on SV1 workstations at NERSC and the T90 vector processor at NPACI.) Second,
the Hybrid algorithm suffers from systematic errors proportional to the discrete time
step used in integrating its stochastic differential equations forward in Monte Carlo
time \cite{HMD}. Most of the simulations reported here used a time step $dt=0.005$ to control
these errors. Future, accurate simulations on larger lattices will need even smaller
time steps. $dt=0.005$  proved adequate here but systematic errors are under
investigation and will be reported on at a later date.

In Fig.2 we show the diquark condensate plotted against the coupling $\beta$. The quark
mass $m$ was set at $0.05$, the coupling was $\beta=1.0$ and $\lambda$ has been extrapolated
to zero using raw data at $\lambda=0.0025$ and $0.005$.

\begin{figure}
\centerline{
\epsfxsize 5 in
\epsfysize 3 in
\epsfbox{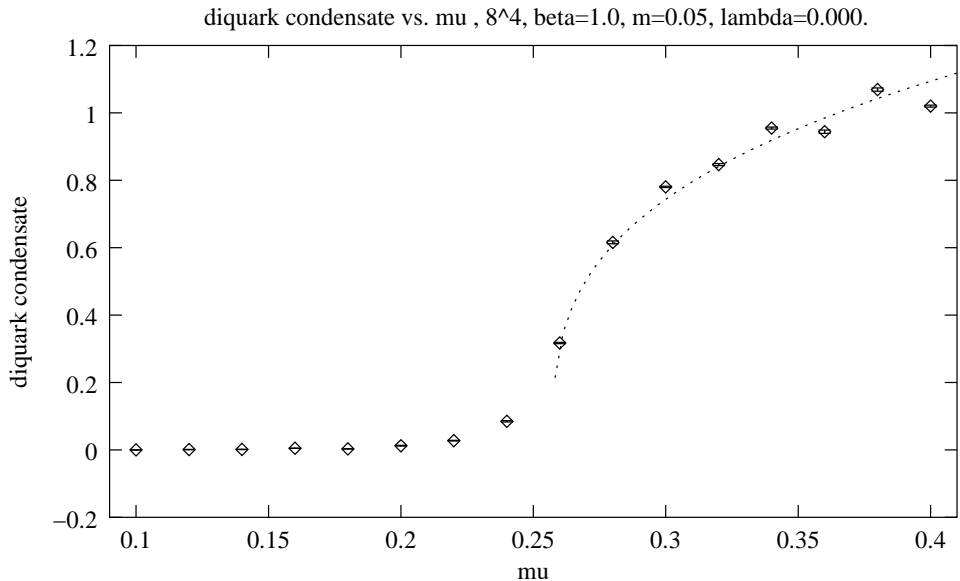}
}
\caption{Diquark Condensate vs. $\mu$}
\end{figure}

We see good evidence to a quark-number violating second order phase transition
in this figure. The dashed line is a power law fit which picks out the
critical chemical potential of $\mu_c = 0.2573(2)$.
Since the coupling here is too large to reside in the gauge theory's scaling window
and the lattice is too small to escape significant coarse-grain errors,
we expect only semi-quanitative results. However, the power law fit is fair, its confidence
level is $7$ percent, and its critical index is $\beta_{mag}= 0.32(5)$ which is somewhat smaller
than the  mean field result $\beta_{mag}=1/2$, predicted by lowest order chiral
perturbation theory \cite{Toublan}, \cite {SUNY}. Note that the quark mass is
fixed at $m=0.05$ throughout this simulation, so chiral symmetry is explicitly broken
and this transition is due to quark number breaking alone. In fact, the chiral order parameter
$\langle\bar{\chi}\chi\rangle$ is nonzero and varies smoothly from $1.00$ to $0.50$ over the
critical region shown in Fig.2. The fermion number density, shown in Fig.3, also shows the
diquark continuous phase transition. The approximate linear dependence of the fermion
number density with $\mu$ is the expected scaling behavior above the transition in lowest order
chiral perturbation theory \cite{Toublan}, \cite {SUNY}.

\begin{figure}
\centerline{
\epsfxsize 5 in
\epsfysize 3 in
\epsfbox{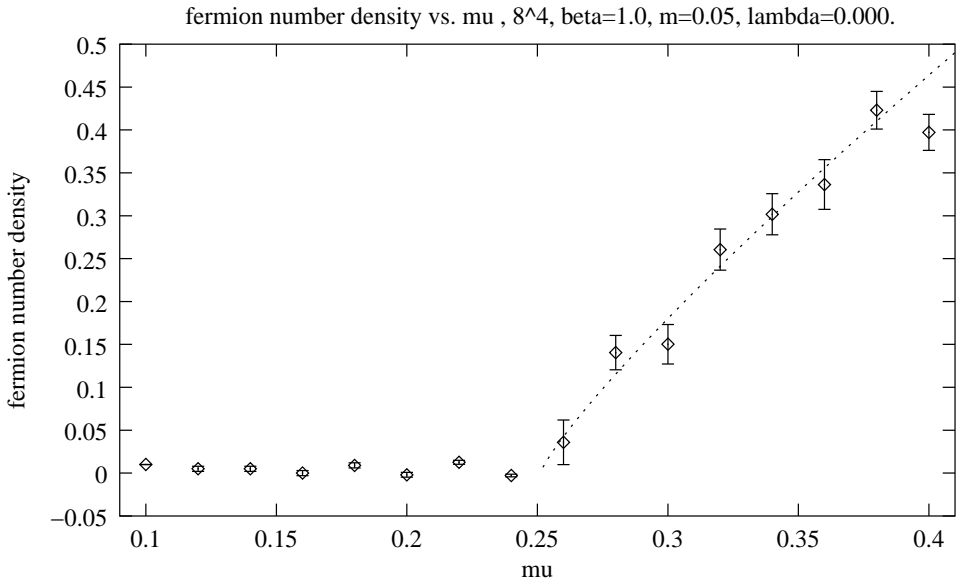}
}
\caption[]{Fermion Number Density vs. $\mu$}
\end{figure}

Now we turn to the main focus of this letter, the phase diagram in the temperature-chemical
potential plane. To begin, consider the small ($8^3 \times 4$) lattice results.

Consider a slice through the phase diagram at a fixed small temperature, for variable $\mu$.
In Fig.4 we show the diquark condensate for $\beta=1.3$, $m=0.05$, with the data extrapolated
to $\lambda=0$ as discussed above. As $\mu$ increases we find that a second order phase transition to
a diquark condensate appears at $\mu_c = 0.2919(4)$. The dashed line fit in Fig.4 has the critical
index $\beta_{mag}=0.50(15)$, in good agreement with mean field theory. The fit has a
confidence level of $21$ percent. The diquark transition
is also clear in a plot of the fermion number density, as expected. However,
the chiral condensate varies smoothly over the
diquark critical region, with  $\langle\bar{\chi}\chi\rangle$ ranging from $0.80$ to $0.50$. The
Wilson Line is also smooth, varying from $0.10$ to $0.20$, with no obvious signs
of a transition.

\begin{figure}
\centerline{
\epsfxsize 5 in
\epsfysize 3 in
\epsfbox{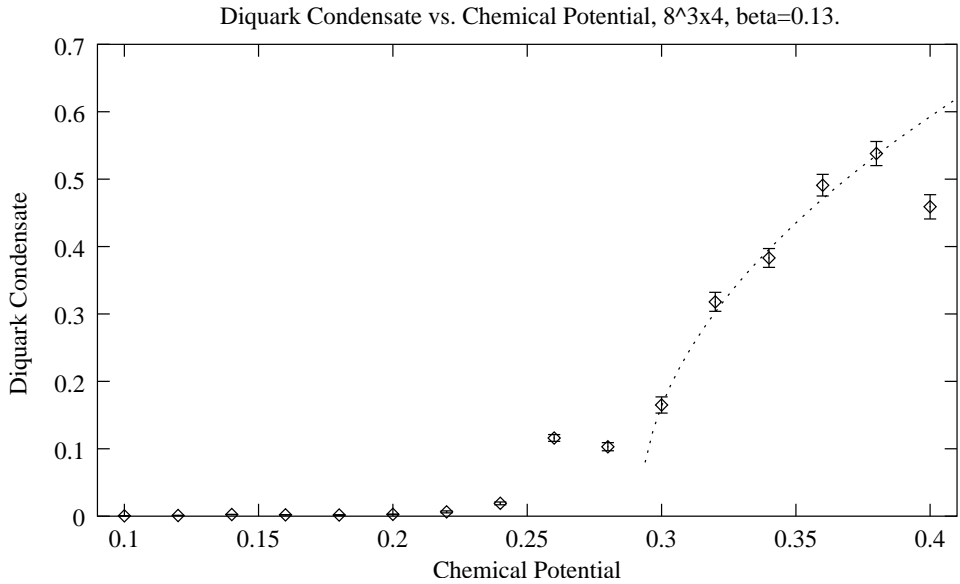}
}
\caption{Diquark Condensate vs. $\mu$.}
\end{figure}

Simulations at $\beta=1.1$ and $1.5$ gave similar conclusions. At $\beta=1.1$, there
was a continuous diquark transition at $\mu=0.2544(3)$ and at $\beta=1.5$ there
was a continuous diquark transition at $\mu=0.2950(3)$. So, as the system is heated ($\beta$
increases), a larger chemical potential is needed to order the system into a diquark condensate.

Next, simulations were run at fixed $\mu$ and variable $\beta$. First, consider a relatively large
and interesting $\mu=0.40$. We show the diquark condensate as a function of $\beta$ for $\mu=0.40$
in Fig.5. There is a clear jump in the condensate at $\beta=1.55(5)$, suggesting a first
order phase transition.

\begin{figure}
\centerline{
\epsfxsize 5 in
\epsfysize 3 in
\epsfbox{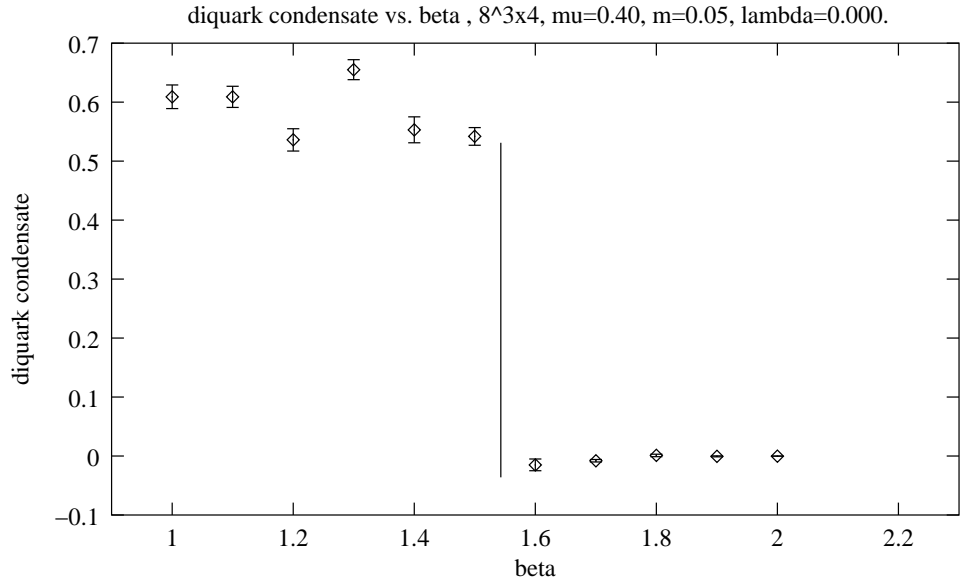}
}
\caption[]{Diquark Condensate vs. $\beta$ for $\mu=0.40$}
\end{figure}

A discontinuous transition is also strongly suggested by the behavior of the Wilson Line
shown in Fig.6.

\begin{figure}
\centerline{
\epsfxsize 5 in
\epsfysize 3 in
\epsfbox{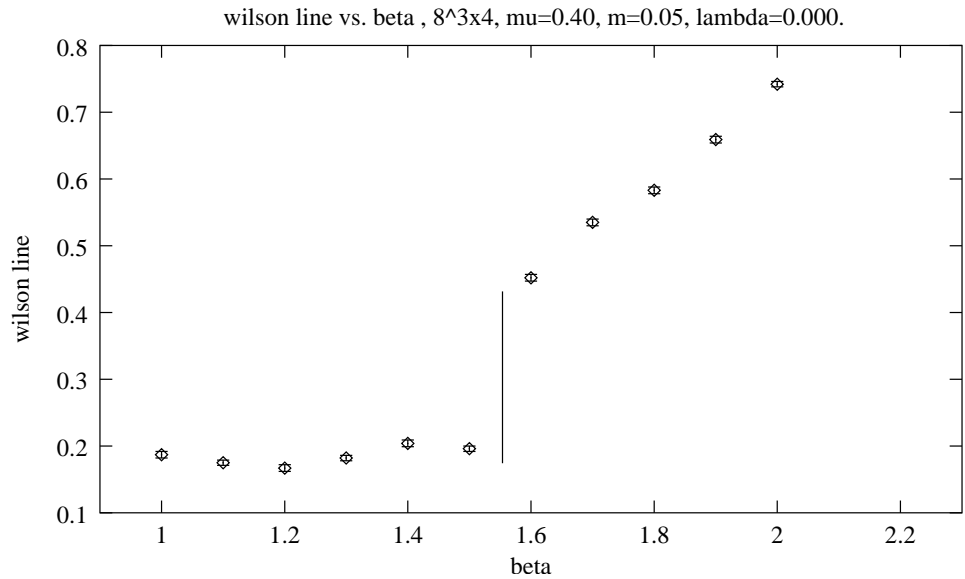}
}
\caption[]{Wilson Line vs. $\beta$ for $\mu=0.40$}
\end{figure}

If this transition at relatively large $\mu$ and $T$ is really first order, then there should be
a tricritical point separating the region of continuous transitions at
smaller $\mu$ and $T$. We strengthened the evidence for a first order transition at
$\mu=0.40$ by running the algorithm on a larger ($12^3 \times 6$) lattice. In Fig.7, 8, and 9
we show the diquark condensate, the Wilson line and the chiral condensate
on the $12^3 \times 6$ lattice.

\begin{figure}
\centerline{
\epsfxsize 5 in
\epsfysize 3 in
\epsfbox{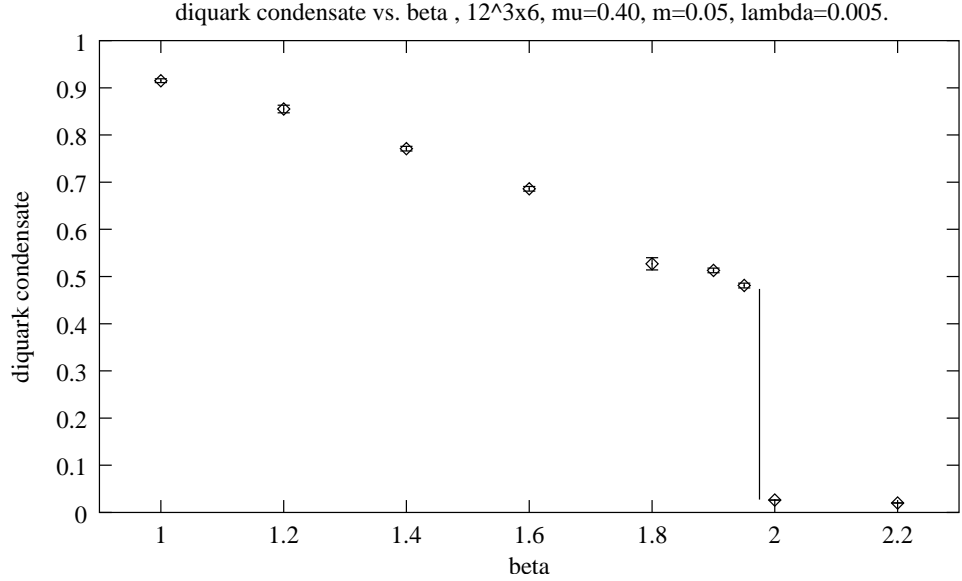}
}
\caption[]{Diquark Condensate vs. $\beta$.}
\end{figure}

\begin{figure}
\centerline{
\epsfxsize 5 in
\epsfysize 3 in
\epsfbox{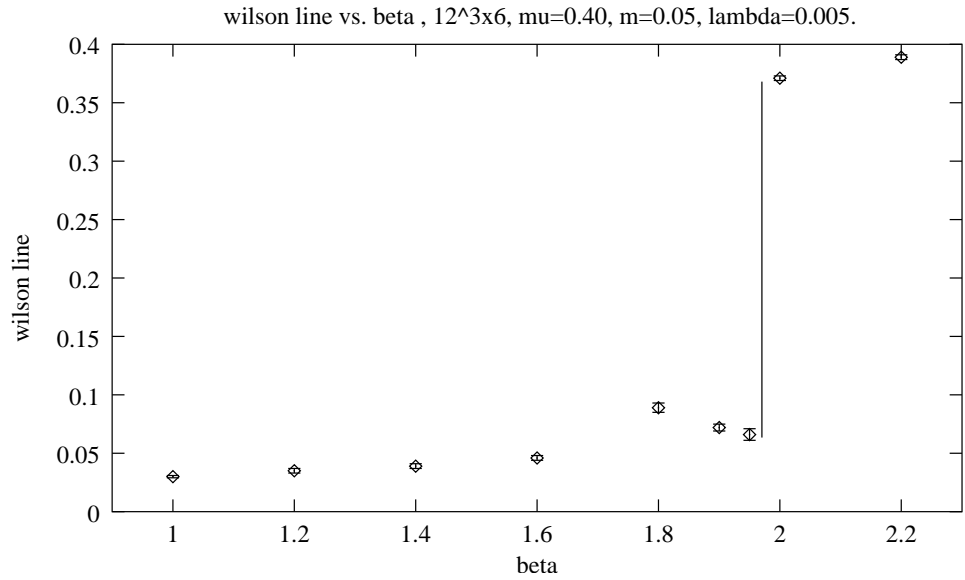}
}
\caption[]{Wilson Line vs. $\beta$.}
\end{figure}

\begin{figure}
\centerline{
\epsfxsize 5 in
\epsfysize 3 in
\epsfbox{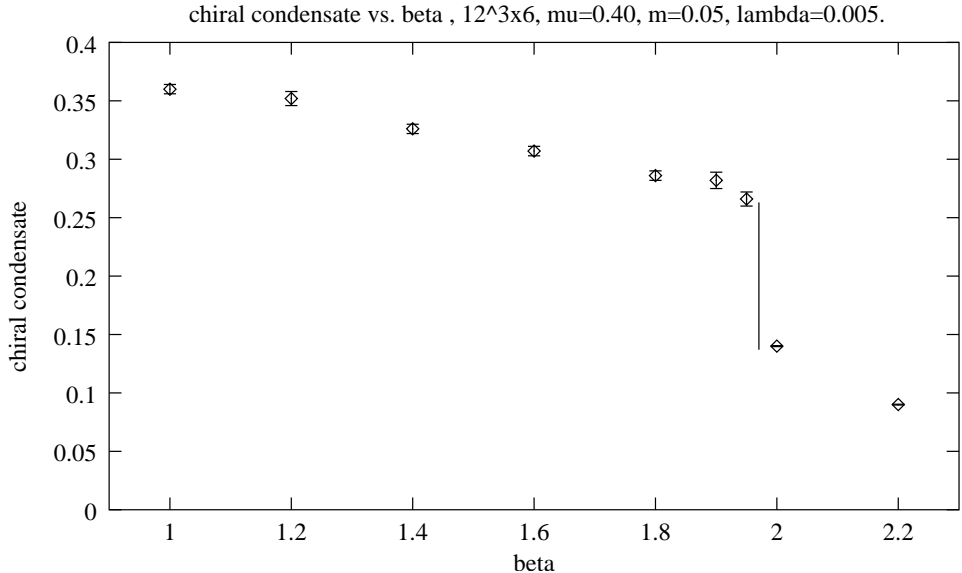}
}
\caption[]{Chiral Condensate vs. $\beta$.}
\end{figure}

In all figures we see evidence for a discontinuous transition at $\beta=1.97(2)$ separating a
diquark condensate and a quark-gluon plasma phase. Note that these are figures of
raw data are taken at $\lambda=0.005$. However, since the transition appears first order,
it should extend to finite
(although probably small) $\lambda$.

$\beta=1.97(2)$  lies in the scaling window of the $SU(2)$ lattice gauge theory with four
species of dynamical quarks, so we can compare our $\mu_c = 0.40$ to the theory's
spectroscopic mass scales \cite{Born}. In fact, $m_{\pi}/2 = 0.31(1)$, so finite $T$ effects have
apparently raised $\mu_c$ somewhat from its zero $T$ value, $m_{\pi}/2$. Such effects will be
discussed below in the context of Effective Lagrangians.

The exact location of the tricritical point should be found in larger scale simulations.
On our small $8^3 \times 4$ lattices we have some evidence that it lies
between $\mu=0.30$ and $0.40$. In particular, we repeated the simulations that produced the data in
Fig.'s 5 and 6, but at $\mu=0.30$ and produced Fig.'s 10 and 11.

 \begin{figure}
\centerline{
\epsfxsize 5 in
\epsfysize 3 in
\epsfbox{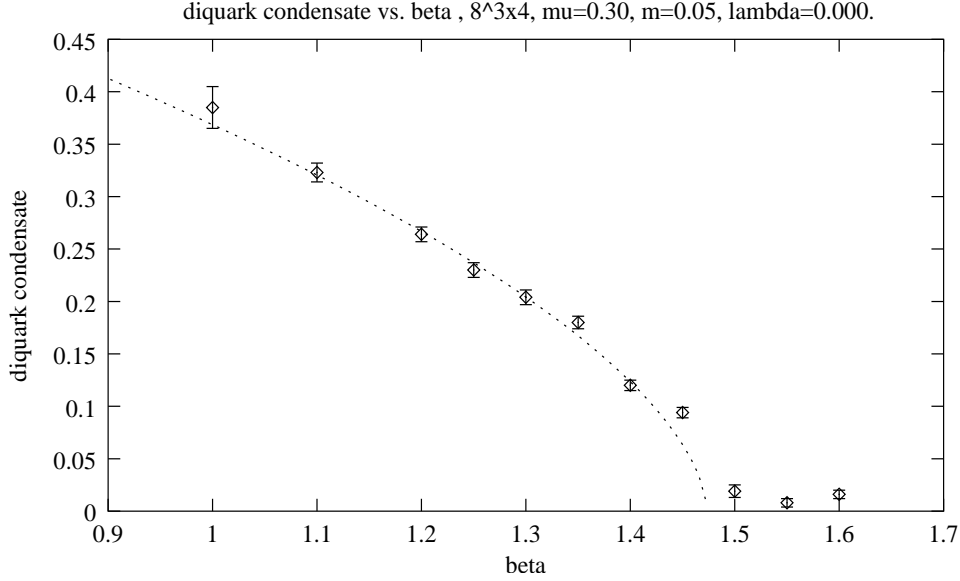}
}
\caption[]{Diquark Condensate vs. $\beta$ for $\mu=0.30$}
\end{figure}

\begin{figure}
\centerline{
\epsfxsize 5 in
\epsfysize 3 in
\epsfbox{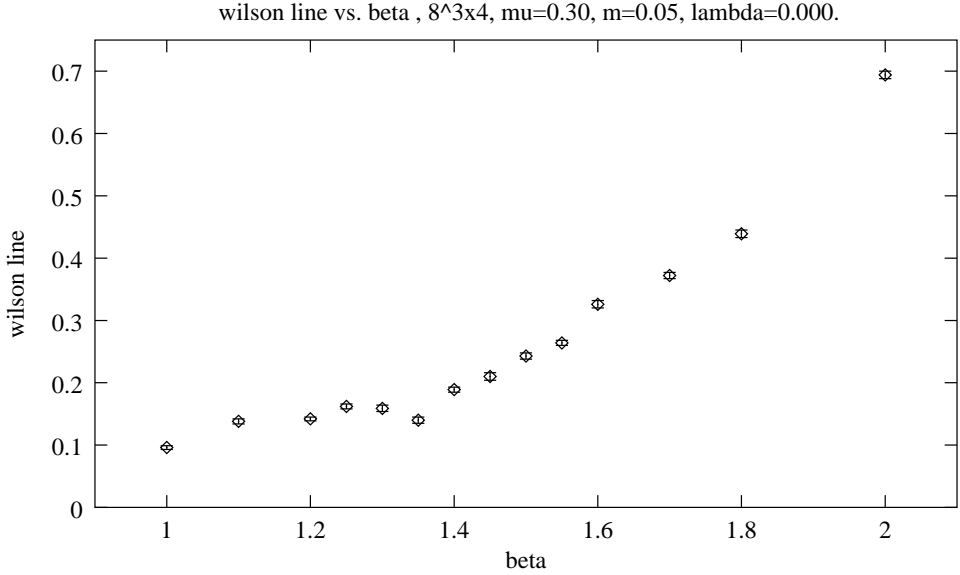}
}
\caption[]{Wilson Line vs. $\beta$ for $\mu=0.30$}
\end{figure}

The dashed curve in Fig.10 is a power law fit to the data which has a confidence
level of $15$ percent, a critical index $\beta_{mag}=0.59(15)$ and critical coupling
$\beta_c= 1.4740(5)$. According to lowest order chiral perturbation theory,
the critical behavior along the entire line below the tricritical point should be in
one Universality Class, Mean Field Theory, while a tricritical point should have
distinct indices, including $\beta_{mag}^{tri}=1/4$. It would be challenging and
particularly interesting to confirm the value $\beta_{mag}^{tri}=1/4$ in the next
round of lattice simulations, on larger lattices closer to the theory's
continuum limit.

Is this tricritical point suggested here expected theoretically?
To answer this question we studied the possibility of a first order
phase transition between the superfluid and the normal phases using
the Landau approach.  Our goal was to analytically understand the
emergence of the first order phase transition observed in the
numerical simulations at high enough $\mu$, not to describe
the second order phase transition for chemical potentials just above
$\mu_c=m_\pi/2$. We
constructed the Landau free energy that describes
the behavior of the diquark condensate for chemical potential greater
than $\mu_c$. Within this approach there
is indeed a first order phase transition that separates the superfluid and
the normal phases for high enough $\mu$.

In order to construct the Landau free energy, we have first to
determine the symmetries of the system \cite{Goldenfeld}.
The low-energy effective Lagrangian that has been used to describe
QCD with two colors and quarks in the fundamental representation at
zero temperature and nonzero chemical potential in \cite{Toublan,SUNY}
is solely based on the symmetries of the QCD partition function.
For small enough temperatures, the symmetries remain the same as at
$T=0$.
Therefore the effective Lagrangian
provides us with a  natural basis to investigate the
symmetries of the system in the superfluid phase at nonzero temperature.
We only need to use the substantial properties of the
low-energy effective Lagrangian. The reader is invited to consult
\cite{Toublan,SUNY} for a more detailed discussion of the
effective Lagrangian itself. The effective Lagrangian
at lowest order reads
\begin{eqnarray}
  \label{Leff}
  {\cal L}_{\rm eff}&=&F_\pi^2 \Big[  \frac12 {\rm Tr} \partial_\nu
  \Sigma^\dagger \partial_\nu \Sigma +2 \mu {\rm Tr} B \Sigma^\dagger
  \partial_0 \Sigma  \\
&&- \mu^2 {\rm Tr} \Big( \Sigma B^T \Sigma^\dagger + BB  \Big)
  - m_\pi^2 {\rm Re} \; {\rm Tr} \hat{M} \Sigma \Big], \nonumber
\end{eqnarray}
where $F_\pi$ is the pion decay constant, and $m_\pi$ is the pion mass.
The matrix $B$
retains the symmetries of the chemical
potential term in the QCD Lagrangian, and the matrix $\hat{M}$ retains the
symmetries of the mass matrix.
Finally the field $\Sigma=U \bar{\Sigma} U^T$
contains both the minimum $\bar{\Sigma}$ of (\ref{Leff}) and the
fluctuations around it, i.e. the
Goldstone fields, $U={\rm exp} (i \pi_a T_a/F_\pi)$ ($T_a$ are the generators
of the Goldstone manifold). At $T=0$ when $\mu$ reaches $\mu_c=m_\pi/2$, the
minimum begins to rotate: $\bar{\Sigma}=\cos \alpha \Sigma_c +
\sin \alpha \Sigma_d$, where $\Sigma_c$ corresponds to the
quark-antiquark condensate, and $\Sigma_d$ corresponds to the diquark
condensate. The angle of rotation is given by $\alpha=0$ for
$\mu<\mu_c$, and by $\cos \alpha=m_\pi^2/4 \mu^2$ for $\mu>\mu_c$
\cite{Toublan,SUNY}. For $\mu>\mu_c$ the ground state of the system is a
diquark condensation phase. The diquark condensate breaks
the symmetry generated by the $U(1)$ baryon charge.

In the superfluid phase ($\mu > \mu_c$) and at zero temperature,
if we expand the effective Lagrangian (\ref{Leff}) in the fluctuations of
$\Sigma$ (i.e. in the Goldstone fields), we find that
there are terms that are even in the Goldstone fields, and that there are
terms that are odd in the Goldstone fields. These odd terms come from
the linear derivative term in (\ref{Leff}); they are absent in the normal
phase.
These odd terms are always present in the superfluid phase, even at small
temperature. They involve the long range excitations that carry the same
quantum number as the diquark condensate.
Hence the Landau free energy can be written as \cite{Goldenfeld}
\begin{eqnarray}
\label{LandauFE1}
  {\tt L}=A t \chi^2 + B \chi^3 + C \chi^4,
\end{eqnarray}
where $\chi$ is the order parameter, and $t=T/T_c-1$ is the reduced
temperature. The order parameter of the superfluid phase is the diquark
condensate.
The Landau free energy contains odd terms. They are not
excluded by the symmetries of the system: as we saw above in
the effective Lagrangian, these symmetries allow the presence of
odd terms in the Goldstone fields that carry the same quantum numbers
as the diquark condensate. This is very different from the situation
at $\mu=0$ where the symmetry excludes odd terms in the
pion fields both for three colors and for two colors
\cite{WilczekP,Wirstam}. From standard studies of
critical phenomena, these odd terms are known to be able to drive a
first order phase transition. We have expanded ${\tt L}$ to $O(\chi^4)$ assuming that
all the essential physics near the critical temperature $T_c$ appears
at this order. Since $\chi=0$ in the symmetric phase above the
critical temperature, there cannot be any linear term in $\chi$,
the coefficient of the quadratic term must be proportional to $t$, and
$A$ has to be positive. Furthermore $C$ has to be positive so that
$|\chi| \rightarrow \infty$ is not a minimum of the Landau free
energy.

Some properties of the coefficients $A$, $B$, and $C$ in
(\ref{LandauFE1}) can be deduced
from the effective Lagrangian (\ref{Leff}). Once we are in
the superfluid phase, the pion mass is practically irrelevant. For
chemical potentials not very far above $\mu_c$, the orientation of the
condensate is almost completely in the diquark direction. From the
effective Lagrangian (\ref{Leff}) expanded to fourth order in the
Goldstone fields and neglecting the pion mass,
we infer that the coefficient in the Landau free energy
(\ref{LandauFE1}) are of the following form
\begin{eqnarray}
  A &=& a \mu^2, \; \; a>0  \nonumber \\
  B &=& b \mu (t+1) \\
  C &=& c \mu^2/F_\pi^2, \; \; c>0. \nonumber
\end{eqnarray}
The temperature dependence of the coefficient of the cubic term $B$
comes from the $\partial_0 \Sigma$ term in the effective Lagrangian
(\ref{Leff}). This coefficient $B$ has to vanish at zero
temperature. Finally we get that the Landau free energy density is
given by
\begin{eqnarray}
  \label{LandauFE}
  {\tt L}=a t  \mu^2 \chi^2+b \mu (t+1) \chi^3
+ c  \frac{\mu^2}{F_\pi^2} \chi^4.
\end{eqnarray}

Since we neglected the pion mass,
the chemical potential we are using here has to be
understood as $\mu-\mu_c=\mu-m_\pi/2$.
The Landau free energy (\ref{LandauFE})
can only be used to describe the transition from the superfluid phase
to the normal phase. Therefore the Landau free energy (\ref{LandauFE})
is only valid for $\mu > \mu_c$.
For $\mu-\mu_c \geq \mu_t=b F_\pi/\sqrt{a  c}$, if $t<(\mu_t/(\mu-\mu_c
+ \sqrt{(\mu-\mu_c)^2-\mu_t^2}))^2$ the minimum is at $\chi \neq 0$,
but for $t>(\mu_t/(\mu-\mu_c
+ \sqrt{(\mu-\mu_c)^2-\mu_t^2}))^2$, the minimum jumps to $\chi=0$.
The Landau free energy (\ref{LandauFE}) therefore predicts a first
order phase transition for $\mu > \mu_c+b
F_\pi/\sqrt{a c}$. For $\mu-\mu_c< \mu_t$ the cubic term cannot
drive a first order transition. From this analysis we thus expect that for
$\mu-m_\pi/2$ of the order of $F_\pi$ a first
order phase transition separates the superfluid phase from
the normal phase.

At $T=0$ a second order phase transition separates the superfluid and
the normal phases. From continuity arguments, the phase transition
between these two phase has to remain second order for small enough
temperatures. With the arguments given above, we find that the phase transition
is going to become first order above some chemical potential. We therefore
find that there is a tricritical point in the phase diagram. Our analysis
is too crude to
tell the exact position of the  tricritical point, but it predicts that it
should correspond to $\mu-m_\pi/2$ of the order of $F_\pi$.

This work will be followed up by several additional investigations. We certainly need to
improve the numerical data by running on larger lattices, closer to the theory's
continuum limit. Once that is done and more spectroscopy data is accumulated, we will be able
to state our results in physical units and check many of the predictions of lowest order chiral
perturbation theory quantitatively. It will also be interesting to study QCD with small
chemical potentials associated with the three light quark flavors \cite{SS}, \cite{TK}.
In particular, we expect the phase diagram at nonzero isospin chemical potential
and temperature to be very similar to the one discussed here.
Although we cannot
attack the $SU(3)$ theory with a large Baryon number chemical potential, these other situations
can be studied both analytically and numerically, and interesting new phases of matter have
been found there which should be investigated further.

This work was partially supported by NSF under grant NSF-PHY96-05199
and by the U.S. Department of Energy under contract W-31-109-ENG-38.
D.T. is supported in part by ``Holderbank''-Stiftung.
The simulations were done at NPACI and NERSC.
B. Vanderheyden, J. Verbaarschot are acknowledged for useful discussions.


\end{document}